\documentclass[conference]{IEEEtran}
\pdfoutput=1
\ifCLASSINFOpdf
\else
\fi

\usepackage{booktabs}
\usepackage{enumitem}
\usepackage{varwidth}
\usepackage{capt-of}
\usepackage{listings}
\usepackage{amsmath}
\usepackage{framed}
\usepackage{tabu}
\usepackage{float}
\usepackage[caption = false]{subfig}
\usepackage{enumitem}
\usepackage{color}
\usepackage{graphicx}
\usepackage{cleveref}
\usepackage{fancybox}
\usepackage[export]{adjustbox}
\usepackage[table]{xcolor}
\definecolor{Gray}{gray}{0.9}
\definecolor{White}{RGB}{255,255,255}
\usepackage{bbold}
\usepackage{bbm}
\usepackage{mathbbol}
\usepackage{multirow}
\usepackage{fix-cm}
\usepackage{siunitx,booktabs}
\usepackage{xcolor}

\usepackage{setspace}
\usepackage{graphicx}
\usepackage{colortbl}
\usepackage{array}
\usepackage{pifont}
\usepackage {rotating}
\usepackage{mwe}
\usepackage{pbox}
\usepackage{enumitem}
\let\oldding\ding% Store old \ding in \oldding
\renewcommand{\ding}[2][1]{\scalebox{#1}{\oldding{#2}}}
\usepackage{cite}

\newcommand{\Csharp}{%
  {\settoheight{\dimen0}{C}C\kern-.05em \resizebox{!}{\dimen0}{\raisebox{\depth}{\#}}}}

 \usepackage{enumitem}
\usepackage{xspace}
\usepackage{xparse}
\DeclareDocumentCommand\newstep{o}{%
\item\IfNoValueTF{#1}{}{#1 \textendash\xspace}}

\newlist{steps}{enumerate}{1}
\setlist[steps]{label=\textit{Step \arabic*:},leftmargin=*}

\definecolor{orange}{RGB}{0,32,96}
\definecolor{o}{RGB}{245,245,245}
\definecolor{g}{RGB}{50,50,50}

\PassOptionsToPackage{gray}{xcolor}
\usepackage{tikz}

\usepackage{tcolorbox}

\begin{document}
\title{\huge A Visual Narrative Path from Switching to Resuming a Requirements Engineering Task}

\author{\IEEEauthorblockN{Zahra Shakeri Hossein Abad, Alex Shymka, Jenny Le, Noor Hammad, Guenther Ruhe}
\IEEEauthorblockA{Department of Computer Science\\ University of Calgary, Calgary, Canada\\
Email: \{zshakeri, alex.shymka, Jenny.le, nour.hammad, ruhe\}@ucalgary.ca}
}


%Telephone: (800) 555--1212, Fax: (888) 555--1212}



\maketitle

\begin{abstract}
Requirements Engineering (RE) is closely tied to other development activities and is at the heart and foundation of every software development process. This makes RE the most data and communication intensive activity compared to other development tasks. The highly demanding communication makes task switching and interruptions inevitable in RE activities. While task switching often allows us to perform tasks effectively, it imposes a cognitive load and can be detrimental to the primary task, particularly in complex tasks as the ones typical for RE activities.
 Visualization mechanisms enhanced with analytical methods and interaction techniques help software developers obtain a better cognitive understanding of the complexity of RE decisions, leading to timelier and higher quality decisions. 
 In this paper, we propose to apply interactive visual analytics techniques for managing requirements decisions from various perspectives, including stakeholders communication, RE task switching, and interruptions. We propose a new layered visualization framework that supports the analytical reasoning process of task switching. This framework consists of both data analysis and visualization layers. The visual layers offer interactive knowledge visualization components for managing task interruption decisions at different stages of an interruption (i.e. before, during, and after). The analytical layers provide narrative knowledge about the consequences of task switching decisions and help requirements engineers to recall their reasoning process and decisions upon resuming a task. Moreover, we surveyed 53 software developers to test our visual prototype and to explore more required features for the visual and analytical layers of our framework.  \end{abstract}


\IEEEpeerreviewmaketitle

\section{Introduction and Motivation}

Task interruptions are a type of task switching or sequential multitasking \cite{Mind}. During the term of performing a task, professional software developers need to frequently interrupt their unfinished tasks for various reasons, such as answering unexpected questions from their coworkers, to address new re-prioritized requirements, or to attend a scheduled team meeting \cite{SERIP}. According to the results of our recent retrospective analysis on \(5,079\) recorded tasks of \(19\) professional software developers, we found that in all of the cases that the disruptiveness of Requirements Engineering (RE) interruptions is statistically different from other software development tasks, RE related tasks are more vulnerable to interruptions compared to other task types \cite{RE17}. This is mainly due to the high level of cognition in RE activities \cite{RO}. 
\begin{figure}
    \centering
    \vspace{-2mm}
    \includegraphics[width=0.47\textwidth]{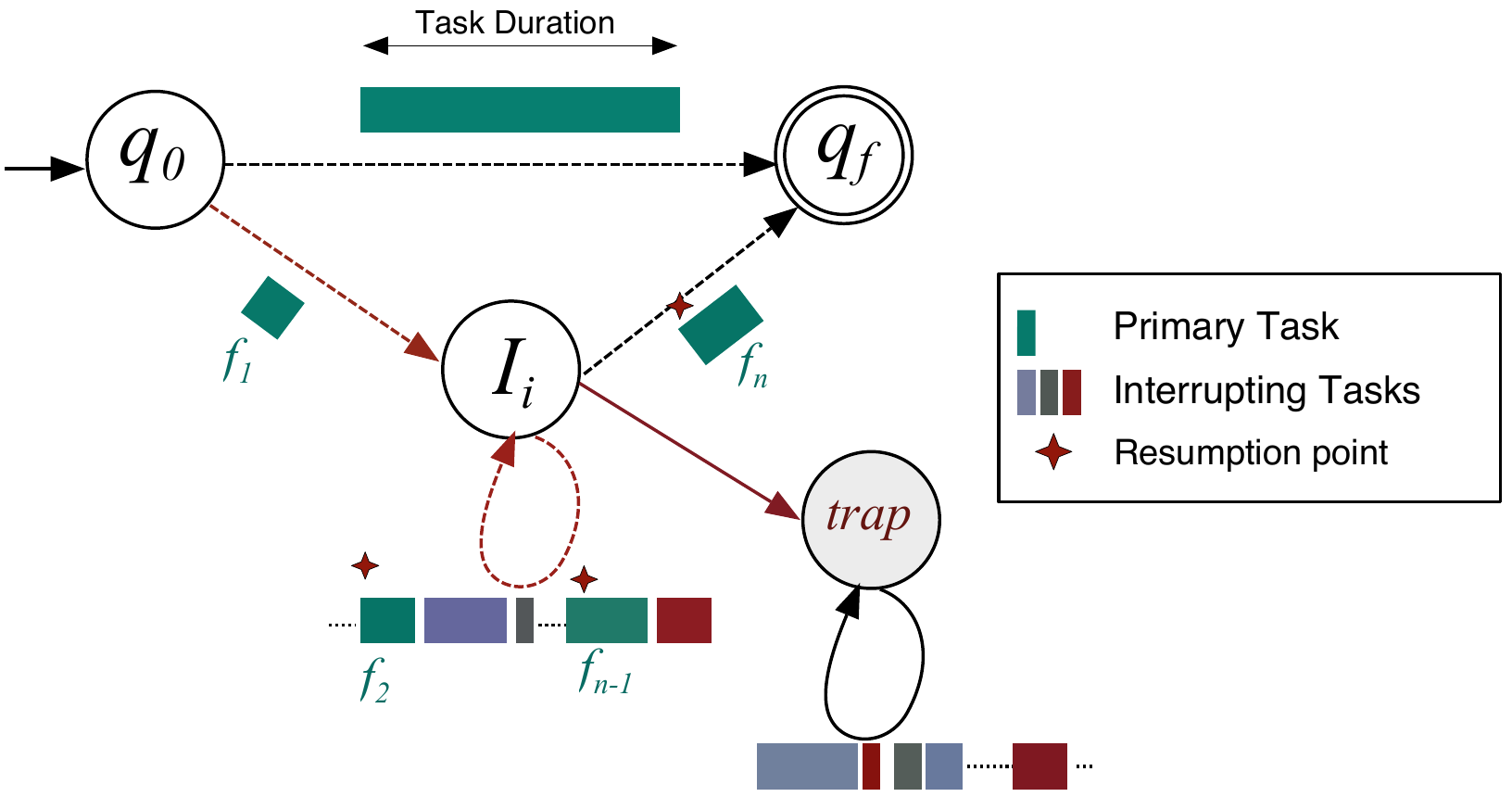}
   \caption{A state diagram for a typical task execution{ [}\(q_0\) and \(q_f\): initial and final states, \(f\)= task fragments, and  \(I_i\)= the \(i^{th}\) interruption{]} }
    \label{fig:state}
    \vspace{-5mm}
\end{figure}
As illustrated in Figure \ref{fig:state}, to execute a typical software development task, the subject might take the direct route to proceed from the initial state of a task (\(q_0\)) to the final state (\(q_f\)). Using the direct link, without any detours on the way, is only one possible way of doing it. However, this is an unrealistic view of the way software developers perform their tasks. The execution of the task might be interrupted by other tasks. Node \(I_i\) represents these interruptions and the interruption loop on this node shows the nested interruptions that might occur during the term of executing the primary task. Considering the limited cognitive flexibility of humans \cite{ICSE}, nested task switchings causes mental congestion for keeping track of multiple states of tasks, which decays the goal of the primary task \cite{behind}. Thus, the greater the value of {\it i}, the more disruptive the interruptions. In our recent study on exploring and understanding tasks interruption in RE activities \cite{RE17}, we found that for 66\% of RE interruptions users did not resume the task right away, and 11\% of interrupted RE tasks never got resumed (i.e. the trap state on Figure \ref{fig:state}). 

In recent years, much research effort has been directed towards visualizing different aspects (e.g. activities, artifacts) of RE \cite{Garm, Irit,  ParisaZahra}.  However, to the best of our knowledge, there is no previous work on the application of visual analytics approaches for reducing the cognitive cost of interruptions and task switching in RE activities. This paper reports on a proposed visual analytics framework, which provides a visual narrative solution for managing interruption decisions and supports the analytical reasoning process of the primary task's resumption. This layered framework consists of three analytical and three visual layers and all of these layers are interrelated. The analytical layers aim to analyze historical interrupted tasks and their associated artifacts and provide additional insights into the visual layers. Moreover, these layers will be used to analyze interactions' histories in terms of various RE artifacts and provide an analytical reasoning process for RE task resumptions. The visual layers offer interactive visualizations which cover the main steps (i.e. before, during, and after) of an interruption. We provide background information about our study in Section \ref{sec:BG}, followed by our research goals and research questions (Section \ref{sec:RQ}). Our proposed visual framework, including our overall research approach and our progress on each layer of this framework, is discussed in Section \ref{sec:Framework}. We conclude the paper by discussing the main contributions and research implications in Section \ref{sec:Expected}.

\section{Background and Related Work}
\label{sec:BG}
\subsection{Terminologies (Task Interruptions in RE)}
\label{sec:Term}
In this section, we briefly summarize the main terminologies that we use in this paper. 

{\bf Definition 1:} The {\it disruptiveness} of an interruption refers to the negative impact it poses on developers' productivity and can be measured in terms of the number of task fragments resulted from this interruption (\(D_1\)), and the length of resumption (\(D_2\)) and interruption (\(D_3\)) lags   \cite{behind, Parnin}.

{\bf Definition 2:} We identified a set of {\it interruption characteristics} in \cite{RE17}, which will be used as the key data points of our proposed visualization methods. A task's context (i.e. project), type, granularity level, progress status, and priority, as well as the timing of interruptions, are some examples of these features. 

{\bf Definition 3:} In our recent study, we found that in the context of RE task switching and interruption, {\it self-interruptions} (i.e. interruptions initiated by the subject) are more disruptive. In this paper, we use this concept to design one of the visual components of our proposed visual framework. 
\subsection{RE Visualization Techniques}
A classification of existing RE visualization techniques based on the visualization type they address (e.g. data, information, or knowledge visualization) and the aspects of RE they cover (e.g. RE activities, Stakeholders, and domain) is presented in our recent Systematic Literature Review (SLR) of RE visualization \cite{Zahra}. In this SLR, we proposed the common RE visualization patterns in the form of \(\bf {\big<content, focus\big>}\), where \({\bf content}\) shows each of the RE activities, and \({\bf focus}\) denotes the \textit{ What, How, Why, Where,} and \textit{Who} components of each visualization technique. The results of this SLR revealed that there is a clear need for more investigation and research to support knowledge visualization in the area of RE. Moreover,  among all RE activities, requirements communication and evolution, as well as Non-Functional Requirements (NFR's) and requirements uncertainties need more investigation.

\begin{figure*}[!bt]
\vspace{-1mm}
\begin{framed}
\vspace{-3mm}
    \centering
    \includegraphics[scale=1.45]{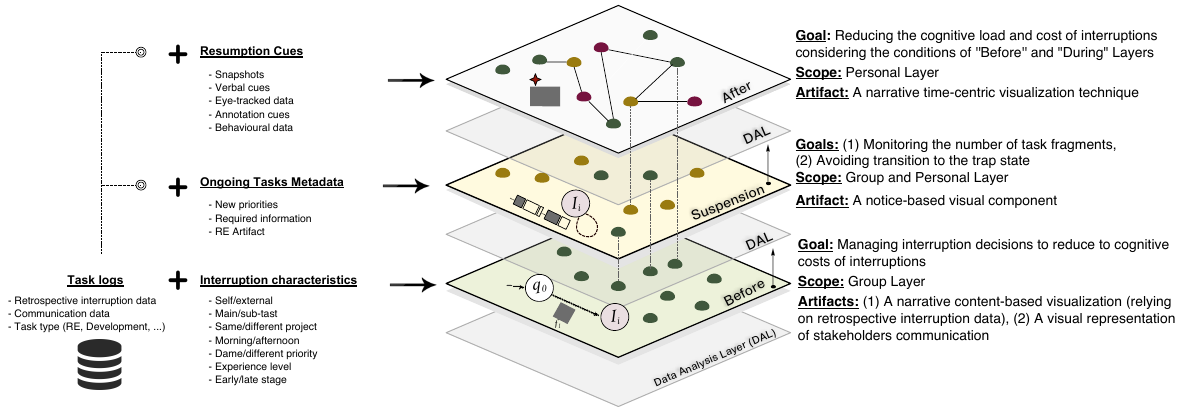}
     \vspace{-7mm}
    \end{framed}
     \vspace{-3mm}
   \caption{The Proposed Interruption Visualization Framework}
    \label{fig:Frame}
    \vspace{-5mm}
    
\end{figure*}

\subsection{Interruption Visualization}

Parnin et al. \cite{Parnin} explore various strategies and coping mechanisms that programmers utilize in order to manage interruptions, while also proposing suggestions on how task resumption can be better supported. A system for handling interruptions is described as having three phases; suspension, resolution and resumption. During the suspension phase, programmers have the choice to preserve their working state by internalization (i.e using memory training to remember the working state) and externalization (i.e applying physical or electronic tools to preserve the working state). Internalization and externalization can be utilized in three suspension strategies: rehearsal, serialization, and cue priming. In the resolution phase,  programmers use various tactics to restore their previous working state. Some restoration techniques are: global restoration, goal restoration, and plan restoration. Lastly, resumption is the last step the programmer would go through before resuming their primary task. Some popular resumption tactics are: return to last stopping place; review task assignment; execute program; restore from task breakdown; and review source code change history. Different programmers would perform the resumption strategy that they are more familiar with. 

Liu et al. \cite{ResVis} conducted an experiment where visual feedback was meant to motivate users to return to the primary task. The visual feedback was either a progress bar that depleted or a flower that withered as the user strayed from the primary task. The task was editing a Microsoft Word document for 40 minutes with no end state. They encouraged participants to go about this task at their leisure, allowing any habits they had such as listening to music or checking emails. Participants with visual feedback spent less time on other tabs and more time editing the word document than the control group with no visual feedback. 

While the existing research provided a wealth of insight on RE visualization and resumption strategies, to the best of our knowledge, there has been no research in the RE community that investigates a visual aid for reducing the cognitive cost of RE task switchings and interruptions and to help requirements engineers' recall and reconstruct their reasoning process.

\section {Goals and Research Questions}

%||||||||||||||||||||||||||||||||||||||||||||||||||||||||||||||||||
\label{sec:RQ}

We intend to develop a narrative visual framework to aid requirements engineers in making swift decisions regarding their task switchings without dealing with complex visualization methods or, worse yet, to deal with the data directly. To this end, we raised the following research questions:

\begin{itemize}
\item {\bf RQ1 [Before Interruption]:} What are effective visualization techniques for making an efficient RE task switching decision (the \(q_0 \rightarrow I_i\) transition in Figure \ref{fig:state})?
\item {\bf RQ2 [Suspension Period]:} What are visualization techniques to monitor the number of task fragments resulted from interruptions (the self-loop on state \(I_i\) in Figure \ref{fig:state})?
\item  {\bf RQ3 [After Interruption]:} What are effective cues and visualization techniques for resuming tasks with a less cognitive cost (all sections marked with \ding {70} in Figure \ref{fig:state})?
\end{itemize}

\section{A Layered Visualization Framework}
\label{sec:Framework}
%----------------------------
An overview of our research approach is modelled in the form of a layered visualization framework and is presented in Figure \ref{fig:Frame}. This section elaborates and discusses a detailed description of each layer of this framework.
\subsection{Data Analysis Layers}
\label{sec:analysis}
 In this section we describe the overall data analysis associated with each {\it ``Data Analysis Layer (DAL)''} of our proposed framework. 
 \subsubsection {Mining Interruption Patterns} 
 To visualize the {\bf before interruption} and {\bf suspension} layers, we apply and customize the Apriori algorithm \cite{Apriori} to mine association rules and explore {\it  disruptiveness patterns}. We define a {\it disruptiveness Pattern} as \(D= \big< (T_{k}, \alpha_{i}) , D_{1-j}\big>\), where \(T_{k}\) represents the type of an RE task (e.g. requirements gathering, analysis, evolution, or validation); \(\alpha_i\) denotes interruption characteristics, such as contextual, temporal, and type of the interruptions; and \(D_j\) represents the disruptiveness factors (Section \ref{sec:Term}), such as interruption lag, resumption lag and the number of fragments resulted from each RE interruption. 
 
 An {\it association rule} \((Z \Rightarrow Y)\) for a pattern \(D= \big< (T_{k}, \alpha_{i}) , D_{j}\big>\) consist of two non-empty sets: \vspace{-1mm}\[Z= \bigcup_{k=1}^{\#types}\bigcup_{j=1}^{n}(T_{k}, \alpha_i) \quad \text{and} \quad Y=\{D_1, D_2, D_3\}.\vspace{-2mm}\]
 These rules are interpreted as:  {``whenever an RE activity {\it k} gets interrupted with interruption characteristics \(\{\alpha_1, \alpha_2, ..., \alpha_n\}\), we have specific disruptiveness measures \(\{D_1, D_2, D_3\}\)''.} The {\it support}  of a {\it disruptiveness pattern} is the percentage of records which contain all parameters in \(Z\cup Y\) and can be used as: \[\text{\it support} = \dfrac{freq(Z,Y)}{\mid Item set\mid}\] 
 
 The {\it confidence} of an association rule \((Y \Rightarrow Z)\) indicates the strength of the rule and can be determined as conditional probablity \(P(Z\mid Y)\) that the disruptiveness value \(Z\) occurred, given the interruption condition \(Y\) already happened.   According to the Apriori algorithm, a pattern occurs frequently if its support is above the \(\text{min}_ {support}\), and an association rule is strong if its confidence is above the \(\text{min}_ {confidence}\) values.
\begin{figure*}[!bt]

    \centering
    \vspace{-3mm}
    \includegraphics[scale=.73]{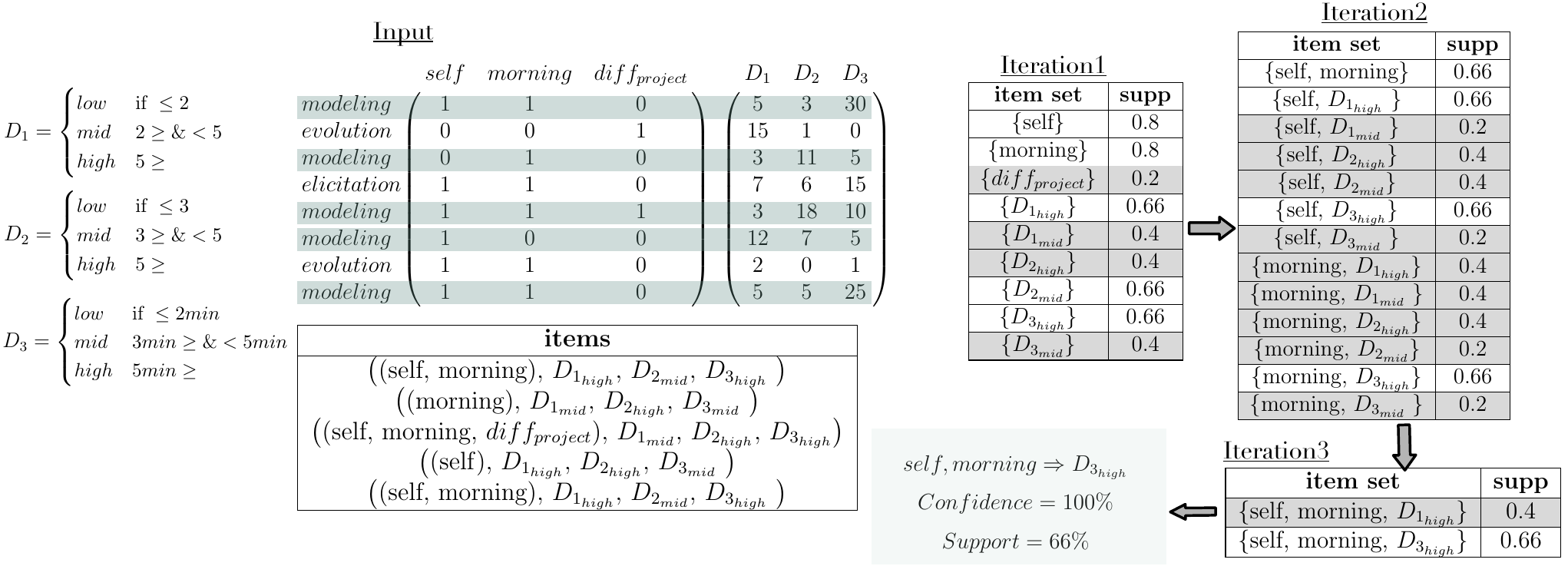}

   \caption{An illustrative case study of the application of association rule mining for layers 1 and 2 }
    \label{fig:Example}
    \vspace{-3mm}
    
\end{figure*}

{\small\ding{228} }{\bf An illustrative case study:} We recorded switched and interrupted tasks of a real world on-going software development project. For the sake of simplicity, we only chose five random ``requirements modeling'' tasks and ran our case study with only three interruption characteristics. The task-characteristics matrix, as the main input of the process, is illustrated in Figure \ref{fig:Example}. The Apriori algorithm takes a \(\text{\it Min}_ {support}\) and a  \(\text{\it Min}_ {confidence}\) as input parameters. To run this illustrative example, we used the following values:\(
    \small
    \begin{cases}
      \text{\it Min}_ {support} & 0.5\\
      \text{\it Min}_ {confidence} & 0.5 
    \end{cases}
\)
for these parameters.

For a pattern \(D= \big< (T_{k}, \alpha_{i}) , D_{j}\big>\) there exist \(2^{\mid T_{k}\cup {\alpha_i}\cup{D_j}\mid}-1\) association rules. To deal with the complexity of our frequent pattern mining approach, we filtered our item sets as follows:  

\begin{itemize}
\item We only consider one RE task type for running the algorithm (the type of the on-going task). For example, if the user intends to switch a requirement modeling task, we only consider this type of task for exploring frequent patterns, as illustrated in Figure \ref{fig:Example}.

\item We ignore all item sets which only belong to one of the {\it interruption characteristics} or {\it disruptiveness measures} sets. For example, as illustrated in Figure \ref{fig:Example}, iteration 2, we do not have any item set like \{\(D_{1{high}}, D_{3_{high}}\)\}.

\end{itemize}

The algorithm iterates over the set of item sets (i.e. the input box on Figure \ref{fig:Example})  and forms
patterns from the interruption characteristics and disruptiveness measures in the same set. We expanded patterns in each iteration based on the {\it support} parameter. The Iterative process continues until all possible extensions are reached (e.g. the \({3^{rd}}\) iteration in this example). The final set of frequent patterns of our study contains only one frequent pattern: {\it ``Self-switching a requirements modeling task in the morning contributes to a greater interruption lag''}, with {confidence of 100\% and support 66\%}. The frequent patterns resulted from this layer will be used as an input for the first and the second narrative visualization layers (Figure \ref{fig:Frame}).

 \subsubsection{Analysis of Interaction Histories}
 \label{sec:Interaction}
\begin{figure*}
\vspace{-6mm}
\centering
\subfloat[The interaction history graph]{\includegraphics[scale=0.4]{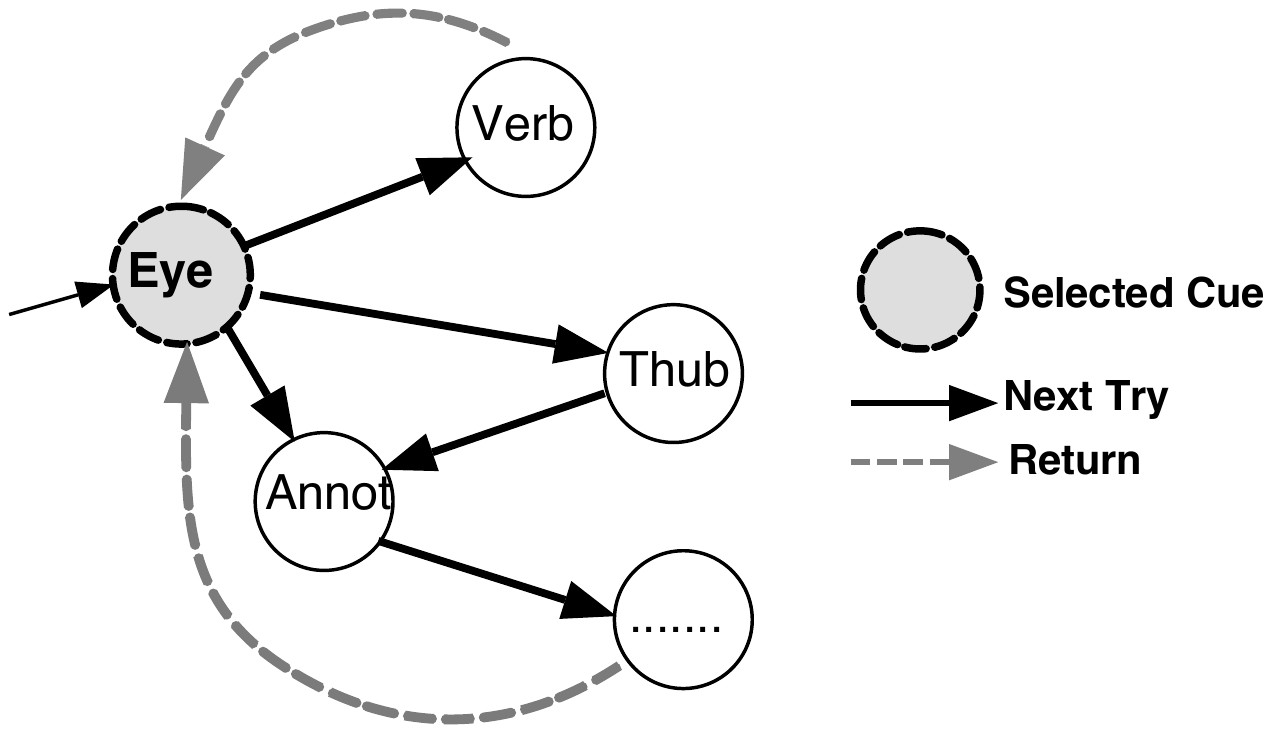}}
\subfloat[Force Directed Layout]{\includegraphics[scale=0.13]{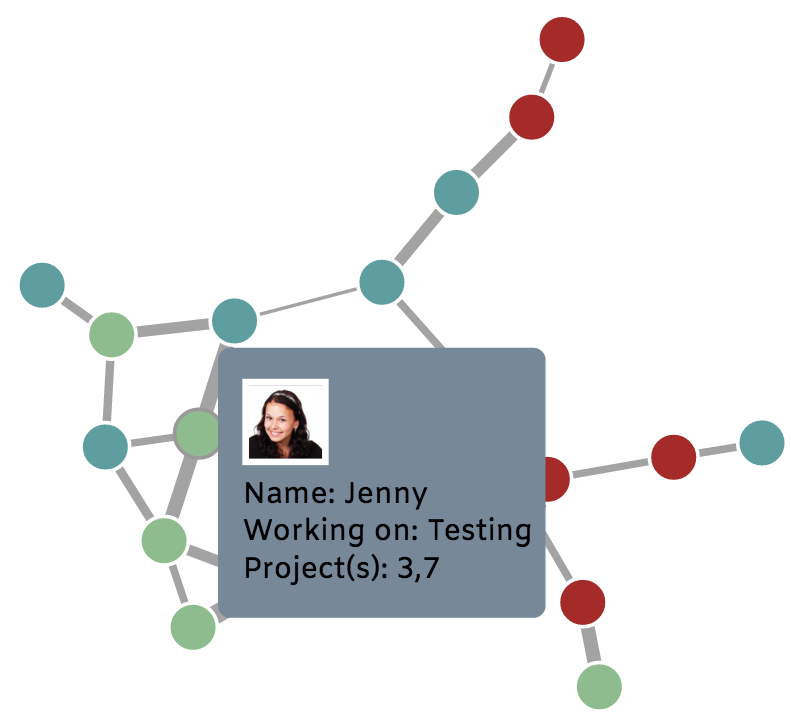}}
\subfloat[Radial Layout]{\includegraphics[scale=0.1]{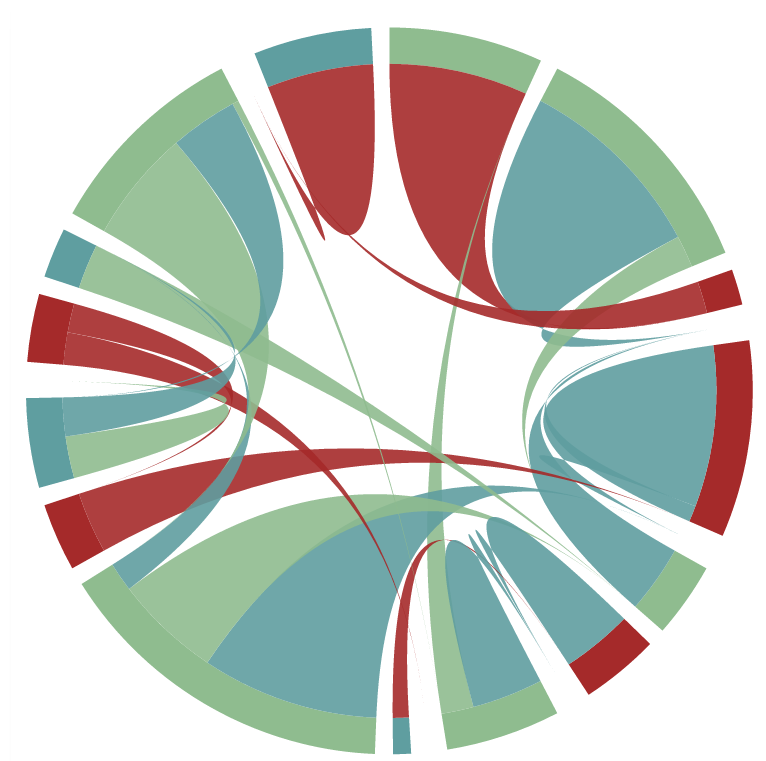}}\hfill
\subfloat[Sankey Layout]{\includegraphics[scale=0.11]{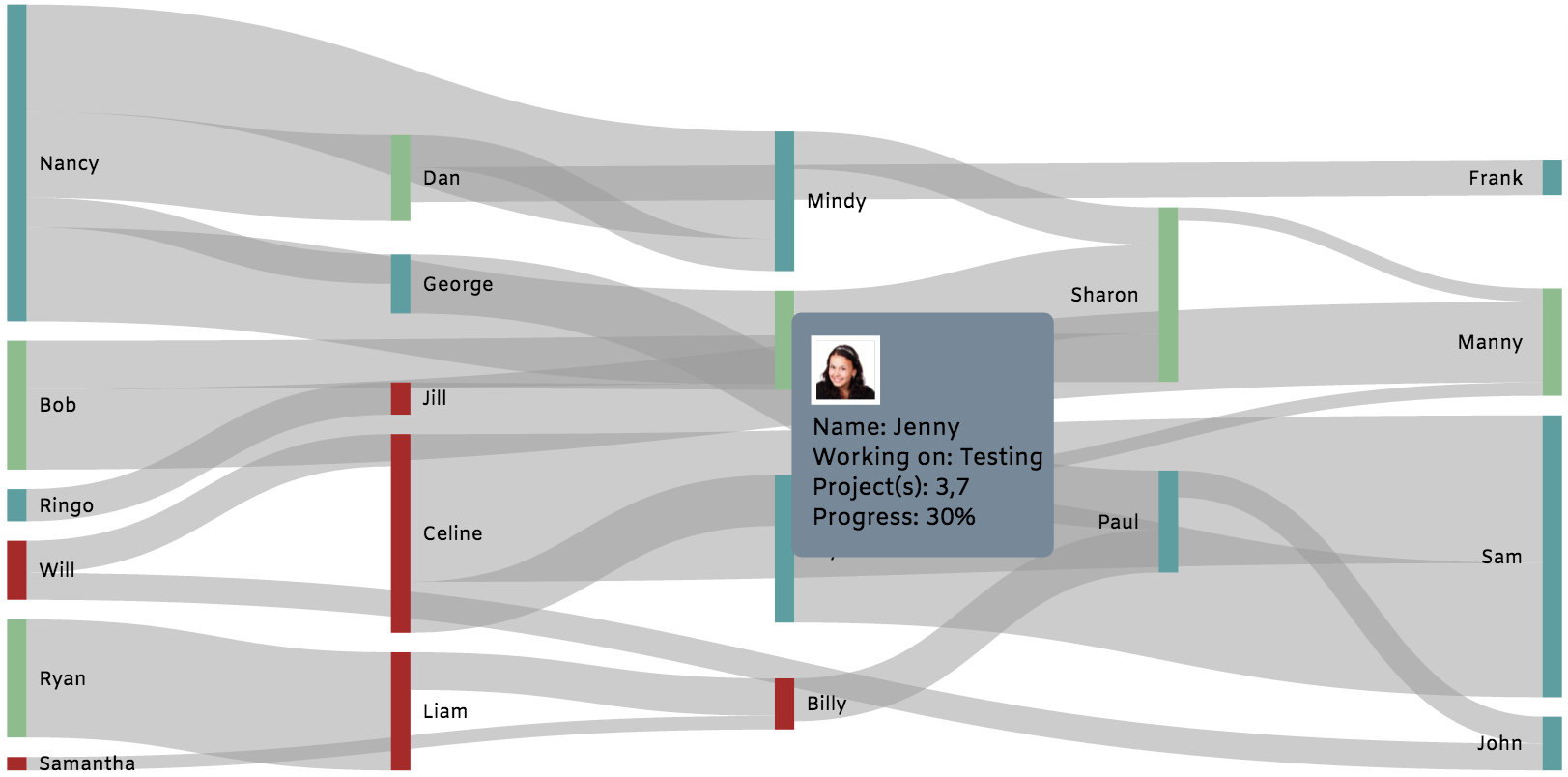}}\hfill
%\subfloat[Sankey Layout]{\includegraphics[scale=0.5]{Figures/Timeline}}
\caption{Analytical and visual components of different layers of the proposed visual framework}
\label{fig:VisExample}
\vspace{-6mm}
\end{figure*}

The design decisions for the third layer (i.e. the resumption layer) are impacted by analyzing the data collected from user interaction logs. To model these histories we use a directional graph of resumption states. Each resumption strategy (e.g. verbal cues, eye-movement data, thumbnail images) is represented as a single node and edges model the sequence of interactions with each of these cues. To discover time-ordered sequences of cues that have been followed by past users, who have been in the same situation, we use the Sequential Association Mining (SAM) approach, as detailed in \cite{SAR}. We use the following format to define each item set: \(\big<(cue_1, t_{1, 1}), (cue_2, t_{2,1}), ..., (cue_m, t_{i,k})\big>\), where {\it i} represents the order of using cues and {\it k} represents the number of times a user navigates to a cue. For example, a possible item set for the sample graph presented in Figure \ref{fig:VisExample} (a) can be defined as:  \(\big<(Eye, t_{1, 1}), (Verb, t_{2,1}), (Eye, t_{3, 2})\big>\).  Each SAM has a degree of support and confidence associated with it. The support of a rule is defined as the fraction of strings in the
set of interaction strings, where the rule successfully applies. We use these patterns to visualize the third layer of our visualization framework.

\subsection{Visualization Layers}
\begin{figure}
\vspace{-2mm}
\centering
\subfloat[ The usage of interruption characteristics]{\includegraphics[scale=0.41]{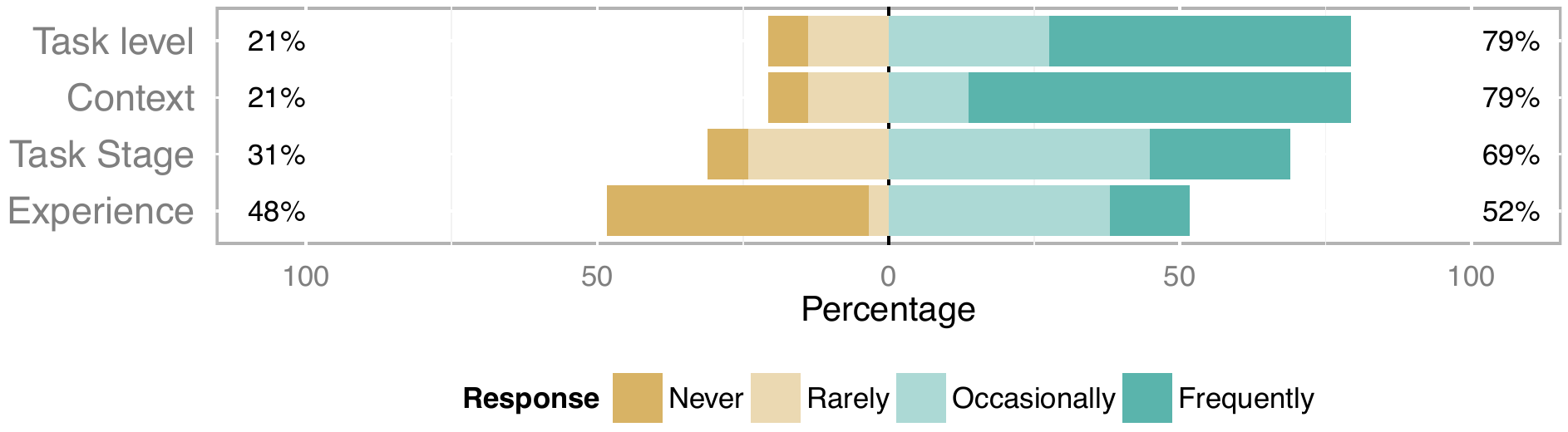}}\\
 \vspace{-3mm}
\subfloat[Testing the usability of prototyped visualization techniques]{\includegraphics[scale=0.41]{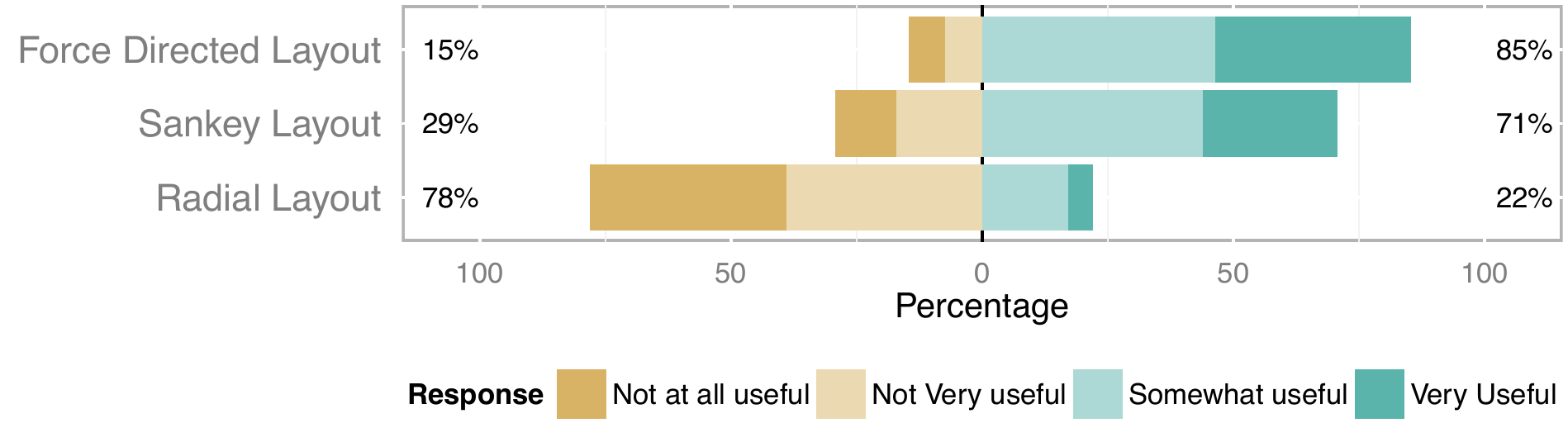}}
\caption{Survey results for required data points on the graphs and usability of the communication graphs}
\label{fig:Likert}
 \vspace{-6mm}
\end{figure}

The visual layers of the proposed framework (Figure \ref{fig:Frame}) consist of visual narrative representations of the main transitions of interruption graph, illustrated in Figure \ref{fig:state}: 
(1) Layer 1 (before interruption) models the \(q_0\rightarrow I_i\) transition, (2) Layer 2 (suspension Layer) provides a visual analytics component to monitor the self-loop on state \(I_i\), and (3) Layer 3 (after interruption) addresses the resumption times, annotated by \ding{70}, and aims to reduce the cognitive cost of interruptions. 

{\small\ding{228} }{\bf User Survey: }To gain more insight into the required feature for each layer and to test the usability of our developed prototype, we surveyed 53 software developers. We used Survey Monkey\footnote{\scriptsize http://www.surveymonkey.com} to design the online survey and collect responses. To recruit our survey participants, we used snowball and random sampling (e.g. LinkedIn and Reddit Forums) methods \cite{Snow}. The survey consists of 10 questions including multiple choice, Likert scale, and open-ended questions. In addition to these questions, we included a direct link to the developed prototype in our survey and asked our participants to interact with each visualization technique and rate them based on their usability and usefulness in representing stakeholders' communication\footnote{https://wcm.ucalgary.ca/zshakeri/files/zshakeri/visualization-survey.pdf}. The average software development experience of participants was {4.5} (\(\pm3\)), and the average of their teamwork experience was {3.5} (\(\pm3\)) years.  This study has been approved by the University of Calgary Conjoint Faculties Research Ethics Board (CFREB  \footnote{\scriptsize http://www.ucalgary.ca/research/researchers/ethics-compliance/cfreb}).

\subsubsection{Visual Layer 1- Before Interruption}

We refer to this layer, as detailed in Figure \ref{fig:Frame}, as a group layer, which expresses the collaborative aspects of task switchings. This layer consists of the fundamental dimensions of a visualization; {\it data, information}, and {\it knowledge} visualization as follows:

 {\ding{192} }{\bf Data Visualization:} This dimension covers graphical representation of unprocessed information including; (1) all influential factors on the disruptiveness of interruptions, such as type, priority, stage, context (i.e. project) and level of all on-going tasks, as well the experience level of task performers. The majority of our survey participants use these parameters before switching their tasks (Figure \ref{fig:Likert} (a)); (2) main triggers of self-interruptions such as boredom and task blockage. To explore the data points of the second category, in addition to our main survey, we asked an online open-ended question from 23 professional software developers about the main reasons of switching their RE tasks. 7 (30\%) stated that they often switch their RE tasks when they get blocked and cannot proceed with the primary task. Lack of information about the current requirements, waiting for stakeholders' feedback, and conflicting requirements are the common reasons for task blockage, as stated by our participants. 6 (26\%) described the ``re-prioritization'' of requirements as the main reason of their self-interruptions. ``Getting bored'' and ``personal schedules'' come next, with each of them being 5 (22\%) and 4 (17\%) participants. One participant also stated that {\it ``I usually get distracted by researching a new technology that could help solve the task at hand''}. While the data points related to disruptiveness factors cover the ``re-prioritization'' and ``personal schedules'', we still needed to add boredom and task blockage to this layer to minimize the number of self-interruptions.

To be able to create distinct visual representations for these data points, we proposed to use {\it visual variables}, such as size, shape, orientation, color, value, and texture \cite{VisVar}. However, choosing an appropriate visual variable to represent each aspect of the data is not a straightforward process. A small change in a particular visual variable can significantly affect the performance of a particular task. To represent the {\it boredom} and {\it requirements uncertainties and conflicts} (i.e. task blockage), we used hue, fuzziness, texture, and objects as the main visual variables and asked our survey participants to choose the appropriate variable for each of these data points. 30 (59\%) participants chose the {\it color} (i.e. red) variable and 28 (54\%) chose {\it objects} as the most appropriate variables for representing uncertainty and boredom, respectively. 

 {\ding{193} }{\bf Information Visualization:} To add more insight to the data points listed above, we added the stakeholders' communication graph, which represents task switching frequencies based on tasks' subjects. To this end,  we prototyped the three main visualization techniques for representing communication, such as Force Directed Layout (Figure \ref{fig:VisExample}(b)), Radial Layout (Figure \ref{fig:VisExample}(c)), and Sankey Layout (Figure \ref{fig:VisExample}(d)). The nodes on each layout are people in a communication network, the thicker the line between nodes, the higher the volume of task switching requests. To evaluate the usability of these techniques, we asked our participants to interact with each visualization and rate them based on their usefulness in understanding the communications. As illustrated in Figure \ref{fig:Likert}(b),  45 (85\%) participants chose the "Forced Directed Layout" (Figure \ref{fig:VisExample} (b)) as a useful technique for understanding the state of the other on-going tasks and the behavior of the communication among stakeholders. For example, a participant stated: {\it ``this technique clearly shows the relationships between each node; good labels when hovering over; good overall representation and different colors of the relationships''.}
We also received some feedback for improving this prototype for the next release, as in: {\it ``I see connections between members but I wasn't able to decipher exactly how they are connected or by what projects. Perhaps pre-loading the profile cards rather than activating only when hovered over''.} Moreover, 38 (71\%) participants found the ``Sankey Layout'' (Figure \ref{fig:VisExample} (d)) technique useful, as one participant stated: {\it ``this technique lets me know quickly who I was communicating with''}. However, some participants found this technique confusing and hard to figure out the flow of information effectively. For the next iterations of improving our visual prototype, we will exclude the ``Radial Layout'' technique (Figure \ref{fig:VisExample} (c)) from our proposed visual framework, as 41 (78\%) participants did not find this approach useful, as in: {\it ``no clue what I am looking at''}. However, this technique might be useful for representing the tracking of resumption cues (Layer 3), which needs more investigation in future studies.

 {\ding{194} }{\bf Knowledge Visualization:} The output of the first analytical layer (i.e. the disruptiveness and interaction patterns), explained in Section \ref{sec:analysis}, are used to provide a visual narrative knowledge in this layer.

\subsubsection{Visual Layer 2- Suspension Period} This layer aims to (1) monitor the number of task fragments resulted from each interruption; (2) shorten the resumption lag; and (3) to avoid the trap situation (i.e. the situation where the interruptee never returns to the interrupted task). We asked our participants whether they use any tool or technique to remind them to return to the primary task, and how they would design this reminder. 37 (69\%) participants stated they do not use any technique or tool for managing the resumption lag of their interrupted task. The participants predominantly described the main features of this reminder as follows:

\begin{enumerate}
\item {\bf Notifications [43 (81\%)]:} Pop-ups, verbal notifications, an encouraging email, and sound effects.
\item {\bf Visual pins [10 (19\%)]:} Representing an image of the interrupted task on the screen, and open tabs on an IDE. 
\end{enumerate}

Some participants provided more details about the timing and main features of the notifications and visual pins, as in: {\it ``It should appear when I'm between tasks, or at least at a reasonable pausing point, which might be indicated when I type "git commit" or "git push''} and in: {\it ``It would know the priority of my tasks and remind me at the most convenient time with my schedule.''}.

Moreover, the disruptiveness patterns produced by applying association rule mining (i.e. Apriori algorithm), as described in Section \ref{sec:analysis}, will be used in this layer to add a narration and insight to the visual reminder. 

\subsubsection{Visual Layer 3- Resumption Time}  This layer aims to provide a visual narrative of what users need to reconstruct their memory after resuming a task. Our survey participants reported that, on average, they need 3.2 (\(-3, +12\)) minutes to refresh their memory about what they were doing before getting interrupted. Visualizations of interaction logs, as stated by Lipford et al. \cite{Think}, can serve as an effective memory aid, allowing analysts to recall additional details of their strategies and decisions. Capturing low-level user actions such as mouse
events, eye tracking \cite{Eye}, and keyboard events \cite{Reasoning}, as well as retrospective verbalization \cite{Think}, and annotating task artifacts \cite{Reasoning} are common approaches to cue the resumption process. We asked our survey participants to rank the usefulness of cues, listed in Table \ref{tab:Cuess}, for constructing their memory state after resuming a task. 33 (63\%) and 28 (52\%) participants found ``annotation cues''  and ``thumbnail images'' more useful technique for recalling their reasoning process, compared to other techniques. 

Moreover, the {\it interaction history patterns} produced by applying the sequential association mining technique, will be used at this layer to design the order and the navigation of resumption cues. In addition, the type of an RE task will be considered in producing the interaction patters. For instance, the appropriate resumption cue for a requirements modeling task might not be such useful for resuming a requirements specification task. As the artifacts of these two tasks are different and have different levels of complexity.

\begin{table}
\centering
\scriptsize
\caption{\scriptsize Participants' responses to the usefulness of various resumption cues (1= most useful, 5= least useful) }
\vspace{-2mm}
\label{tab:Cuess}
\begin{tabular}{|c|ccccc|}\hline
{\bf Cues}&1&2&3&4&5\\\hline
Annotation cues & \cellcolor{Gray}46\%&\cellcolor{Gray}17\%&2\%&21\%&12\%\\
Thumbnail images &\cellcolor{Gray}15\%&\cellcolor{Gray} 37\%&26\%&12\%&10\%\\
Verbal cues& 24\%& 12\%&17\%&24\%&22\%\\
Eye cues & 12\%&20\%&22\%&24\%&22\%\\
Behavior graph & 5\%&15\%&29\%&\cellcolor{Gray}17\%&\cellcolor{Gray}34\%\\\hline                                            
\end{tabular}
\vspace{-5mm}
\end{table}

\section{Discussion and Research Implications}
\label{sec:Expected}
The main contribution of this paper lies primary in providing a multi-layered hybrid visualization technique for better managing requirements communication and interruptions. In summary, this research aims at: (1) exploring a novel hybrid visualization technique to reduce the cognitive load of requirements communication and interruptions; (2) providing a new narrative ``knowledge visualization'' technique (i.e. storytelling)  in the area of RE by integrating data analysis and visualization techniques; and (3) proposing a set of time-centric visualization techniques for reducing the cognitive cost of RE task interruptions and classify these techniques based on RE activities and artifacts (e.g. requirements elicitation, communication, and evolution). We addressed our current progress on RQ1-3 in Section \ref{sec:Framework}. Further, we presented an illustrative case study with different interruption characteristics (e.g. self/external, time, and task context) and disruptiveness measures (\(D_{1-3}\)) to clarify our approach for implementing the analytical layers. 

Regarding the evaluation of our proposed approach, we plan to design and run several user studies to empirically evaluate the effectiveness of our approach and the usability of proposed visualization technique. To assess the performance of the analytical approach we use in analytical layers (i.e. association rule mining and sequential association mining), we plan to run our framework on different scales of real world datasets. However, the filtering method we described in Section \ref{sec:analysis} helps reduce the complexity of large-scale datasets. Further, we aim to apply classification techniques \cite{DataTrack} on RE artifacts (e.g. requirements specification) to involve the requirements' type (i.e. functional and non-functional) in reasoning process of the analytical layers. In summary, we believe that this research will foster a clear connection between various areas of research: requirements engineering, data analysis, data visualization, and Human Computer Interaction (HCI).

%||||||||||||||||||||||||||||||||||||||||||||||||||||||||||||||||||

%----------------------------------


\section{Acknowledgement}
This research was supported by the Natural Sciences and Engineering Research Council of Canada (NSERC) Discovery Grant 486565-15 and by an Engage Grant.

%\printbibliography[resetnumbers=true,notcategory=Studies]
\bibliographystyle{IEEEtran}
\bibliography{IEEEabrv,RENext17}

\end{document}